\documentclass{aa}  

\usepackage{graphicx}
\usepackage{txfonts}
\usepackage{algorithm}
\usepackage{algpseudocode}
\usepackage{natbib}
\bibpunct{(}{)}{;}{a}{}{,} 

\newcommand{\tvect}[2]{
  \ensuremath{\Bigg(\negthinspace\begin{matrix}#1\\#2\end{matrix}\Bigg)}}
\newcommand{\minus}{\scalebox{0.75}[1.0]{$-$}}

\begin{document} 

   \title{A novel inversion algorithm for weak gravitational lensing using quasi-conformal geometry}

   \author{Jan Jakob
          \inst{1,*}, Björn Malte Schäfer \inst{1, \#}
          }

   \institute{\inst{1} Zentrum für Astronomie der Universität Heidelberg, Astronomisches Rechen-Institut, Philosophenweg 12, 69120 Heidelberg, Germany\\
              * jan.jakob@stud.uni-heidelberg.de, \textsuperscript{\#} bjoern.malte.schaefer@uni-heidelberg.de
             }

   \date{Preprint published in January 2025}
 
  \abstract
  {The challenge in weak gravitational lensing caused by galaxies and clusters is to infer the projected mass density distribution from gravitational lensing measurements, known as the inversion problem.}
  {We introduce a novel theoretical approach to solving the inversion problem. The cornerstone of the proposed method lies in a complex formalism that describes the lens mapping as a quasi-conformal mapping with the Beltrami coefficient given by the negative of the reduced shear, which can, in principle, be observed from the image ellipticities.}
  {We propose an algorithm called QCLens that is based on this complex formalism. QCLens computes the underlying quasi-conformal mapping using a finite element approach by reducing the problem to two elliptic partial differential equations that solely depend on the reduced shear field.}
  {Experimental results for both the Schwarzschild and the singular isothermal lens demonstrate the agreement of our proposed method with the analytically computable solutions.}
  {}


   \maketitle

\section{Introduction}

Gravitational lensing has become one of the most important fields in present-day astronomy, largely driven by considerable improvements in observational capabilities. Its distinguished feature of being independent of the nature and physical state of the deflecting mass makes it perfectly suited to study dark matter in the Universe. In recent years, interest has increased in exploring the two most dominant components of the Universe: dark matter and dark energy. To this end, large-scale imaging and spectroscopic surveys are currently being undertaken, such as those of the \textit{Euclid} mission (\citealt{EuclidCollaboration2024}), launched in July 2023, the \textit{Rubin} Observatory (Legacy Survey of Space and Time; \citealt{Brough2020}), and the \textit{Roman} Space Telescope (\citealt{Spergel2015}), set to begin in late 2025, which will map the sky with unprecedented accuracy. A prominent cosmological probe for these surveys is weak gravitational lensing.

Weak gravitational lensing refers to the subtle distortions observed in the images of distant galaxies caused by the gravitational influence of massive structures along the line of sight. This phenomenon manifests in two primary ways: (i) a convergence field, $\kappa$, leads to the magnification or demagnification of the background galaxies' images, altering their apparent size and brightness, and\ (ii) the shear, $\gamma$, stretches the galaxies' shapes, causing them to appear more elliptical or skewed than they intrinsically are.

The convergence field, $\kappa$, cannot be observed directly due to the mass-sheet degeneracy \citep{Bartelmann2001,Kilbinger2015}. Physically, it represents the projected total matter density along the line of sight, modulated by a lensing kernel in the mid-distance between the observer and the galaxy sources. A widely used mass-mapping algorithm is the Kaiser-Squires method (\citealt{kaiser1993mapping}), which operates as a simple linear operator in Fourier space. However, this method has limitations, such as not accounting for missing data or the effect of noise.

In this paper we propose using a complex formalism for weak lensing, first introduced by \cite{Straumann1997} to describe the lens mapping as quasi-conformal (q.c.) mapping with a Beltrami coefficient field given by the negative of the reduced shear, which can be deduced from the observed image ellipticities. The resulting q.c. mapping can then be decomposed into two elliptical partial differential equations (PDEs) for each component of the complex deflection angle field. To our knowledge, this is the first time that solving q.c. mappings has been proposed for mass-mapping reconstruction.

This paper is structured as follows. In Sect. \ref{chap:2} we introduce the formalism of weak gravitational lensing and describe the mass-mapping reconstruction problem. We also provide a brief overview of the current Kaiser-Squires algorithm. Section \ref{chap:3} shows that weak lensing corresponds to a q.c. mapping from the image to the source plane. We then present our proposed inversion method in Sect. \ref{chap:4}. The method is novel in the sense that it exploits the fact that the lens mapping is q.c. In Sect. \ref{chap:5} we illustrate the feasibility of the proposed method by comparing the computed solutions with the analytical solutions for both the Schwarzschild and the singular isothermal sphere lens model.

\section{Weak lensing mass-mapping}
\label{chap:2}

Gravitational lensing occurs when light from distant galaxies bends around a foreground mass distribution. This phenomenon distorts the appearance of these galaxies, with the extent of distortion depending on the shape and size of the mass distribution along the line of sight. The relationship between the original source coordinates ($\beta$) and the observed, lensed image coordinates ($\theta$) is described by the lens equation 

\begin{equation}
\label{lens_eq}
    \beta = \theta - \nabla \psi (\theta),
\end{equation}

\noindent where $\psi$ describes the lensing potential (\citealt{Umetsu_2020}). By introducing local Cartesian coordinates $\boldsymbol{\theta} = (\theta_1,\theta_2)$ centered on a certain reference point in the image plane, the Jacobian matrix of the lens mapping describing the local properties of lensing becomes
\begin{equation}
\mathcal{A}(\boldsymbol{\theta}) := \frac{\partial \boldsymbol{\beta}}{\partial \boldsymbol{\theta}} = 
\left(\begin{matrix}
1-\psi_{11} & - \psi_{12}\\
-\psi_{12} & 1-\psi_{22}.
\end{matrix} \right),
\end{equation}
with $\psi_{i,j} = \partial^2 \psi / \partial \theta_i \partial \theta_j$ $(i,j=1,2)$. Alternatively, the components can be written as $\mathcal{A}_{ij} = \delta_{ij} - \psi_{ij}$, where $\delta_{ij}$ denotes the Kronecker delta. It is convenient to decompose $\mathcal{A}$ by means of the Pauli matrices $\sigma_a$ ($a=1,2,3)$ as
\begin{equation}
\mathcal{A} = (1-\kappa)\mathcal{I} - \gamma_1\sigma_3 - \gamma_2\sigma_1.
\end{equation}
The $\kappa$ is called the convergence and is defined as one half of the Laplacian of $\psi$:
\begin{equation}
\label{poisson_eq}
\kappa := \frac{1}{2}\left(\psi_{11} + \psi_{22}\right) = \frac{1}{2}\Delta \psi,
\end{equation}
with $\Delta = \boldsymbol{\nabla}_{\theta}^2$. Here $\gamma_1$ and $\gamma_2$ are the two components of the shear, $\gamma$, which can be considered as a complex quantity, $\gamma(\boldsymbol{\theta}) := \gamma_1(\boldsymbol{\theta}) + \mathrm{i}\gamma_2(\boldsymbol{\theta})$. The $\kappa$, $\gamma_1$, and $\gamma_2$ are linear combinations of the second-order derivatives of $\psi$:
\begin{equation}
\gamma_1 := \frac{1}{2}\left(\psi_{11} - \psi_{22}\right),
\end{equation}
\begin{equation}
\gamma_2 := \frac{1}{2}\left(\psi_{12} + \psi_{21}\right) = \psi_{12}.
\end{equation}
Equation (\ref{poisson_eq}) can be regarded as the two-dimensional Poisson equation
\begin{equation}
\label{posisson_eq_full}
\Delta \psi(\boldsymbol{\theta}) = 2\kappa(\boldsymbol{\theta}),
\end{equation}
with inhomogeneity $2\kappa$. Often, one assumes that the field size is (hypothetically) infinite, i.e., it is sufficiently larger than the characteristic angular scale of the lensing cluster, but small enough for the flat-sky assumption to be valid. Then, the Green function becomes $\Delta^{-1}(\boldsymbol{\theta},\boldsymbol{\theta'}) = \mathrm{ln}|\boldsymbol{\theta},\boldsymbol{\theta'}| / (2\pi)$, which yields $\psi$ as convolution of $\Delta^{-1}$ with $2\kappa$:
\begin{equation}
\psi(\boldsymbol{\theta}) = \frac{1}{\pi}\int \mathrm{ln}(\boldsymbol{\theta}-\boldsymbol{\theta'})\kappa(\boldsymbol{\theta'})\mathrm{d}^2\theta'.
\end{equation}
Using these new quantities, $\mathcal{A}$ can be expressed as\begin{equation}
\label{jac_lens_eq}
\mathcal{A}(\boldsymbol{\theta}) = 
\left(\begin{matrix}
1-\kappa-\gamma_1 & \minus \gamma_2\\
\minus\gamma_2 & 1-\kappa+\gamma_1
\end{matrix} \right).
\end{equation}
In the weak lensing limit ($|\kappa|,|\gamma| \ll 1$) we obtain
\begin{equation}
(\mathcal{A}^{-1})_{ij} \simeq (1+\kappa)\delta_{ij} + \Gamma_{ij} \quad (i,j = 1,2).
\end{equation}
Here, $\Gamma_{ij}$ is the matrix defined by (\citealt{Bartelmann2001})
\begin{equation}
\Gamma_{ij} = \left( \partial_i \partial_j - \delta_{ij} \frac{1}{2} \Delta \right)\psi(\boldsymbol{\theta}).
\end{equation}
Equation (\ref{jac_lens_eq}) illustrates that the convergence causes an isotropic change in the size of the source image, as it appears in the diagonal of the matrix \( \mathcal{A} \). In contrast, the shear causes anisotropic distortions in the image shapes. The convergence, $\kappa$, can also be interpreted via Eq. (\ref{posisson_eq_full}) as a weighted projection of the mass density field between the observer and the source. By factoring out the term \( (1 - \kappa) \) in Eq. (\ref{jac_lens_eq}), the amplification matrix depends only on the reduced shear,
\[
\mathcal{A} = (1 - \kappa) \begin{bmatrix} 
1 -g_1 & -g_2 \\
-g_2 & 1 + g_1 
\end{bmatrix},
\]
which is defined as 
\begin{equation}
g := \frac{\gamma}{1 - \kappa}.
\end{equation} 
$g$ can be directly measured in lensing surveys. For the subcritical regime where $\mathrm{det}\mathcal{A} > 0$ we can observe $g$ directly, whereas for negative-parity regions with $\mathrm{det}\mathcal{A} < 0$ the quantity $1/g^*$ is observable.

In this paper we are interested in recovering the convergence, $\kappa$, from reduced shear data. This inverse problem is ill-posed due to the finite sampling of the reduced shear over a restricted survey area and the presence of shape noise in the measurements. However, in this work, the focus is not on addressing measurement limitations like done by \cite{starck_weak_lensing}; instead, we present a theoretical approach that offers an alternative to the Kaiser-Squires method.

\paragraph{Kaiser-Squires.}Following \cite{Meneghetti2021}, we give a short summary of the Kaiser-Squires inversion algorithm, which belongs to the class of free-form methods. In 1993, Kaiser and Squires developed an algorithm for reconstruction convergence maps from the observed weak lensing shear. Today, this algorithm is widely known as the KS93 algorithm. Since the shear and convergence are both linear combinations of the second-order derivatives of the lensing potential, they can be expressed in Fourier space as
\begin{align}
\tilde{\kappa} &= \minus \frac{1}{2}(k_1^2 + k_2^2)~\tilde{\psi},\\
\tilde{\gamma}_1 &= \minus \frac{1}{2}(k_1^2 - k_2^2)~\tilde{\psi},\\
\tilde{\gamma}_2 &= \minus k_1k_2 \tilde{\psi},
\end{align}
where $\tilde{\cdot}$ denotes the Fourier transform of the corresponding quantity, and $k_1$ and $k_2$ the elements of the wave vector, $k,$ with norm square $k^2 = k_1^2 + k_2^2$. With the three independent equations, we can now eliminate $\psi$ and express $\gamma$ as a function of $\kappa$:
\begin{equation}
\label{gamma_from_kappa}
\tvect{~\tilde{\gamma}_1~}{~\tilde{\gamma}_2~} = k^{-2} \tvect{~k_1^2 - k_2^2~}{~2k_1k_2~}~\tilde{\kappa}= A\tilde{\kappa},
\end{equation}
with the operator
\begin{equation}
A := k^{-2} \tvect{~k_1^2 - k_2^2~}{~2k_1k_2~}.
\end{equation}
$A$ transforms the convergence to the shear vector in Fourier space. Using that $A$ is idempotent ($
AA^T = 1$), inverting Eq. (\ref{gamma_from_kappa}) yields $\kappa$ in dependence of $\gamma$:
\begin{equation}
\tilde{\kappa} = A^T \tvect{~\tilde{\gamma_1}~}{~\tilde{\gamma_2}~}.
\end{equation}
We can transform this relation back to real space by taking the inverse Fourier transform,
\begin{equation}
\label{fourier_integral_kappa}
\kappa(\boldsymbol{\theta}) = \frac{1}{\pi} \int_{\mathbb{R}^2} \Big[D_1(\boldsymbol{\theta} - \boldsymbol{\theta '})\gamma_1(\boldsymbol{\theta '}) + D_2(\boldsymbol{\theta} - \boldsymbol{\theta '})\gamma_2(\boldsymbol{\theta '})     \Big]\mathrm{d}^2 \theta ',
\end{equation}
where $D_1$ and $D_2$ are appropriate kernel functions given by
\begin{align}
D_1(\theta_1,\theta_2) &= \frac{\theta_2^2 - \theta_1^2}{\theta^4},\\
D_2(\theta_1,\theta_2) &= \frac{2\theta_1\theta_2}{\theta^4}.
\end{align}
By defining the complex kernel function
\begin{equation}
D(\boldsymbol{\theta}) = D_1(\boldsymbol{\theta}) + \mathrm{i}D_2(\boldsymbol{\theta}),
\end{equation}
Eq. (\ref{fourier_integral_kappa}) can be written as
\begin{equation}
\kappa(\boldsymbol{\theta}) = \frac{1}{\pi} \int_{\mathbb{R}^2} \mathrm{Re}[D^*(\boldsymbol{\theta} - \boldsymbol{\theta '})\gamma(\boldsymbol{\theta '})]\mathrm{d}^2 \theta '.
\end{equation}
As mentioned by \cite{Seitz1996}, under the assumption of vanishing shear at infinity, partial integration yields
\begin{equation}
\kappa(\boldsymbol{\theta}) = \frac{1}{\pi} \int_{\mathbb{R}^2} \boldsymbol{\mathrm{H}}^{KS}(\boldsymbol{\theta '},\boldsymbol{\theta}) \cdot \tvect{~\gamma_{1,1}(\boldsymbol{\theta '}) + \gamma_{2,2}(\boldsymbol{\theta '})~}{~\gamma_{2,1}(\boldsymbol{\theta '}) - \gamma_{1,2}(\boldsymbol{\theta '})~} \mathrm{d}^2 \theta ',
\end{equation}
with 
\begin{equation}
\boldsymbol{\mathrm{H}}^{KS}(\boldsymbol{\theta '},\boldsymbol{\theta}) = \frac{1}{2\pi}\frac{\boldsymbol{\theta} - \boldsymbol{\theta '}}{|\boldsymbol{\theta} - \boldsymbol{\theta '}|^2} = \nabla_{\boldsymbol{\theta '}}\left( \minus \frac{1}{2\pi} \mathrm{ln}~|\boldsymbol{\theta} - \boldsymbol{\theta '}| \right).
\end{equation}
This means that in this limit, the surface mass density is obtained by convolving the deflection angle field of a point mass with the first derivatives of the shear field.

\section{Quasi-conformal mass mapping}
\label{chap:3}
Following \citet{Straumann1997}, we used the Wirtinger calculus to transform the basic lensing equations into a complex formulation. In particular, we see that weak lensing corresponds to a q.c. mapping from the image to the source plane with the Beltrami coefficient given by the reduced shear field, $g$.

\paragraph{Wirtinger calculus.} By identifying $\mathbb{C}$ with $\mathbb{R}^2$, we can write $z \in \mathbb{C}$ as $z = x + \mathrm{i}y$ for $x,y \in \mathbb{R}$. Let $U$ be an open subset of $\mathbb{C}$. The two 1-forms $\mathrm{d}z = \mathrm{d}x + \mathrm{i}\mathrm{d}y$ and $\mathrm{d}\overline{z} = \mathrm{d}x - \mathrm{i}\mathrm{d}y$ form a corresponding basis of the cotangent space of all points in $U$ ($T_zU \cong \mathbb{C}$ for all $z \in U$). By defining the so-called Wirtinger derivatives,
\begin{equation}
\partial_z = \frac{\partial}{\partial z} := \frac{1}{2}\left(\frac{\partial}{\partial x} - \mathrm{i}\frac{\partial}{\partial y}\right),~~\partial_{\overline{z}} = \frac{\partial}{\partial \overline{z}} := \frac{1}{2}\left(\frac{\partial}{\partial x} + \mathrm{i}\frac{\partial}{\partial y}\right),
\end{equation}
\noindent we can represent the differential of any smooth complex function $f$ on $U$ as
\begin{equation}
\mathrm{d}f = \frac{\partial f}{\partial z}\mathrm{d}z + \frac{\partial f}{\partial \overline{z}} \mathrm{d}\overline{z}.
\label{eq:diff_f}
\end{equation}
\noindent We introduce $f_z$ and $f_{\overline{z}}$ for $\partial_zf$ and $\partial_{\overline{z}}f$, respectively, and denote with $\mathcal{D}(U)$ the $\mathbb{C}$-algebra of all functions $f:~U\rightarrow\mathbb{C}$, which are infinitely often differentiable according to the real coordinates $x$ and $y$. Then, according to the Cauchy-Riemann differential equations, the vector space $\mathcal{O}(U)$ of holomorphic functions on U is equal to the kernel of the mapping, $\partial_{\overline{z}}: \mathcal{D}(U) \rightarrow \mathcal{D}(U)$ (see \citealt{Forster2012}). With the Wirtinger derivatives, the Laplacian can be expressed as
\begin{equation}f
\Delta = 4\partial_z \partial_{\overline{z}}.
\label{eq:laplace_operator}
\end{equation}

\paragraph{Differential of the lens mapping.} By applying this formalism to the basic lens equation $\beta: \mathbb{R}^2 \mapsto \mathbb{R}^2, \theta \rightarrow \beta(\theta)$ in (\ref{lens_eq}), $\beta$ can be written as complex function:
\begin{equation}
f: \mathbb{C} \rightarrow \mathbb{C},~z \mapsto f(z) = z - 2\partial_{\overline{z}} \Psi = \partial_{\overline{z}}(z\overline{z} - 2\Psi).
\label{eq:complex_lens_map}
\end{equation}
Using Eqs. (\ref{eq:laplace_operator}) and (\ref{eq:complex_lens_map}), the Poisson equation, Eq. (\ref{posisson_eq_full}), becomes
\begin{equation}
2\partial_z\partial_{\overline{z}}\Psi = \kappa,
\label{eq:complex_poisson_eq}
\end{equation} 
and similarly, the shear vector becomes
\begin{equation}
\partial_{\overline{z}}^2\Psi = \frac{1}{4}(\partial_1^2 - {\partial_2}^2)\Psi + \frac{\mathrm{i}}{2}\partial_1 \partial_2 \Psi = \frac{1}{2}(\gamma_1 + \mathrm{i}\gamma_2) = \frac{1}{2}\gamma.
\label{eq:complex_shear_relation}
\end{equation}
With Eqs. (\ref{eq:complex_lens_map}), (\ref{eq:complex_poisson_eq}), and (\ref{eq:complex_shear_relation}) we can determine the differential of $f$:
\begin{equation}
\mathrm{d}f = \partial_zf \mathrm{d}z + \partial_{\overline{z}}f\mathrm{d}\overline{z} = (1-\kappa)\mathrm{d}z - 2\partial_{\overline{z}}^2\Psi \mathrm{d}\overline{z} = (1-\kappa)\mathrm{d}z - \gamma \mathrm{d}\overline{z}.
\label{eq:diff_lens_map}
\end{equation}

\paragraph{Beltrami equation and quasi-conformal mappings.} A function $f: \Omega_1 \rightarrow \Omega_2$, which is assumed to be at least continuously partially differentiable, between two domains of the complex plane,  $\Omega_1$ and $\Omega_2$, fulfills the Beltrami equation if
\begin{equation}
\frac{\partial f}{\partial \overline{z}} = \mu \frac{\partial f}{\partial z}
\label{eq:beltrami_eq}
\end{equation}
holds on $\Omega_1$, where $\mu$ is a complex-valued function on $\Omega_1$ and Lebesgue measurable. $\mu$ is called the dilatation or Beltrami coefficient of $f$ and contains all information about the conformality of $f$. The Beltrami equation plays a crucial role in the theory of q.c. mappings: $f$ is said to be q.c. if it fulfills the Beltrami equation, Eq. (\ref{eq:beltrami_eq}), and
\begin{equation}
||\mu||_{\infty} := \underset{x \in U}{\text{ess sup}}~|\mu(x)| \leq k < 1 
\label{eq:mu_ll_1}
\end{equation}
holds for some $k \in \mathbb{R}$. Considering the Jacobian $J_f$ of f given by
\begin{equation}
J_f = |f_z|^2 - |f_{\overline{z}}|^2 = |f_z|^2(1-|\mu|^2).
\end{equation}
It is clear that $f$ is q.c. if it fulfills the Beltrami equation and preserves orientation ($J_f > 0$). Furthermore, $||\mu||_{\infty} = 0$  if and only if $f$ is conformal. Thus, q.c. mappings can be seen as generalizations of conformal mappings. Quasi-conformal  mappings are the homeomorphisms that map infinitesimal circles to ellipses of bounded eccentricity (see \citealt{Ming2013}).

\paragraph{Lens equation as quasi-conformal mapping.} By comparing Eq. (\ref{eq:diff_f}) with Eq. (\ref{eq:diff_lens_map}), we obtain the Beltrami coefficient of the lens mapping as the negative of the reduced shear:
\begin{equation}
\mu = \frac{f_{\overline{z}}}{f_z} = \minus \frac{\gamma}{1-\kappa} = \minus g.
\end{equation}
In the weak lensing limit, where $\kappa, \gamma \ll 1$, the lens mapping not only fulfills the Beltrami equation, but also $|g| \leq k < 1$ is satisfied for some $k \in \mathbb{R}$. Otherwise, the Jacobian $J_f$ would become singular, as in the case of multiple images and strong lensing. The lens equation $f$ can therefore be interpreted as a q.c. mapping, which is uniquely determined by the negative of the reduced shear as the Beltrami coefficient and some appropriate boundary conditions. Only in the case of $||\gamma||_{\infty} = 0$, $f$ reduces to a conformal mapping.\\

\noindent As examples for lens mappings and their corresponding Beltrami coefficients, we consider the Schwarzschild and the singular isothermal lens, two important examples of lenses. For the Schwarzschild lens, the lensing object is treated as a point mass in the lens plane. We obtain
\begin{equation}
f(z) = z - \frac{1}{\overline{z}}, \quad \mu = \frac{1}{\overline{z}^2}
\end{equation} 
for the complex lens mapping and the Beltrami coefficient, and for the singular isothermal lens
\begin{equation}
f(z) = z - \frac{z}{|\overline{z}|}, \quad \mu = \frac{z^2}{2|\overline{z}|^3 - |\overline{z}|^2}.
\end{equation}
In Sect. \ref{chap:4} we compare the results of our proposed method against those of analytically solvable examples to show the validity of our algorithm.

\begin{figure}
    \centering
    \includegraphics[width=0.7\linewidth]{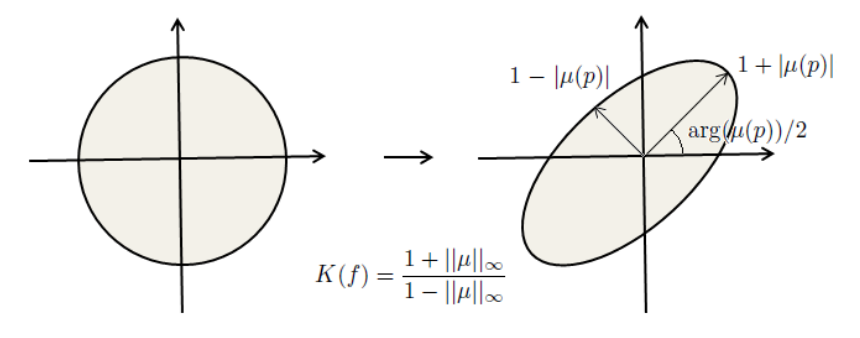}
    \caption{Geometric interpretation of q.c. mappings (figure from \citealt{Ming2013}).}
    \label{fig:geometric_int_beltrami}
\end{figure}

\paragraph{Geometric interpretation.} Let us consider an infinitesimal ellipse field that is constructed in the following way: As shown in Fig. \ref{fig:geometric_int_beltrami}, we assign each point $z \in U$ an infinitesimal circle that is mapped by $f$ to an infinitesimal ellipse of bounded eccentricity:
\begin{equation}
K_f(z) := \frac{|f_z| + |f_{\overline{z}}|}{|f_z| - |f_{\overline{z}}|} = \frac{1 + |\mu(z)|}{1 - |\mu(z)|}.
\label{eq:eccentricity_pointwise}
\end{equation}
The $K_f(z)$ is called the dilatation of $f$ at $z$. By taking the (essential) supremum over all points in $U$, we obtain the notion of the dilatation of $f$:
\begin{equation}
K_f := \underset{z \in U}{\text{ess sup}}~K(f,z) = \frac{1 + ||\mu||_{\infty}}{1 - ||\mu||_{\infty}},
\label{eq:eccentricity_functionwise}
\end{equation}
which is well defined for a q.c. mapping because of $1 - ||\mu||_{\infty} \geq 1 - k > 0$. The argument of the major axis $a = 1 + |\mu(z)|$ of this infinitesimal ellipse can also be expressed in terms of the Beltrami coefficient by
\begin{equation}
\text{arg}(1+|\mu(z)|) = \text{arg}(\mu(z))/2.
\label{eq:argument_ellipse}
\end{equation}

\noindent Geometrically, this means that there is a fixed bound in the stretching for $f$ in any given direction compared to any other direction. Solving the Beltrami equation, Eq. (\ref{eq:beltrami_eq}), is then equivalent to finding a function, $f$, whose associated ellipse field (with bounded eccentricity) coincides with the prescribed Beltrami coefficient field, $\mu$. This is just the inversion problem in gravitational lensing, where the negative of the reduced shear, $g,$ takes over the role of $\mu$.

\section{Modeling with quasi-conformal mappings}
\label{chap:4}

\subsection{Reduction to elliptic PDEs}
The Beltrami equation, Eq. (\ref{eq:beltrami_eq}), can be reduced to two elliptic PDEs for the real and imaginary part of $f$ with coefficients determined by the Beltrami coefficient field $\mu$ (\citealt{Lui2013}). By decomposing $\mu$ and $f$ into $\mu = \text{Re}(\mu) + \mathrm{i}\text{Im}(\mu) =: \rho + \mathrm{i}\tau$ and $f = \text{Re}(f) + \mathrm{i}\text{Im}(f) =: u + \mathrm{i}v$, the Beltrami coefficient can be written in terms of $x$ and $y$ derivatives of $u$ and $v$ as
\begin{equation}
\mu = \rho + \mathrm{i}\tau = \frac{(v_x-v_y) + \mathrm{i}(v_x + u_y)}{(u_x+v_y) + \mathrm{i}(v_x-u_y)}.
\end{equation}
The $v_x$ and $v_y$ can be expressed as linear combinations of $u_x$ and $u_y$,
\begin{equation}
\minus v_x = \alpha_1u_x + \alpha_2u_y,
\label{eq:v_x_lin_comb}
\end{equation}
\begin{equation}
v_y = \alpha_2u_x + \alpha_3u_y,
\label{eq:v_y_lin_comb}
\end{equation}
with
\begin{equation}
\alpha_1 = \frac{(\rho-1)^2 + \tau^2}{1-\rho^2-\tau^2};~\alpha_2 = \minus \frac{2\tau}{1-\rho^2-\tau^2};~\alpha_3 = \frac{(1+\rho)^2+\tau^2}{1-\rho^2-\tau^2}.
\end{equation}
On the other hand,
\begin{equation}
u_y = \alpha_1v_x + \alpha_2v_y,
\end{equation}
\begin{equation}
\minus u_x = \alpha_2v_x + \alpha_3v_y.
\end{equation}
Due to the symmetry of the second derivatives, it holds that
\begin{equation}
\nabla \cdot \tvect{~\minus v_y~}{~v_x~} = 0 \quad\text{and}\quad \nabla \cdot \tvect{~u_y~}{~\minus u_x~} = 0.
\label{eq:symm_2nd_deriv}
\end{equation}
By substituting Eqs. (\ref{eq:v_x_lin_comb}) and (\ref{eq:v_y_lin_comb}) into Eq. (\ref{eq:symm_2nd_deriv}), we obtain two elliptic PDEs for $u$ and $v$,
\begin{equation}
\nabla \cdot \Bigg(~A~\tvect{~u_x~}{~u_y~}~\Bigg) = 0~\text{and}~\nabla \cdot \Bigg(~A~\tvect{~v_x~}{~v_y~}~\Bigg) = 0,
\label{eq:elliptic_pdes}
\end{equation}
where the symmetric, positive definite matrix, $A,$ is given by
\begin{equation}
A = \Bigg(\begin{matrix}
  \alpha_1 & \alpha_2\\
  \alpha_2 & \alpha_3
\end{matrix}\Bigg).
\label{eq:matrix_A}
\end{equation}
The eigenvalues $\lambda_1$ and $\lambda_2$ of $A$,  
\begin{equation}
\lambda_1 = (1 - |\mu|)^2,
\end{equation}
\begin{equation}
\lambda_2 = (1 + |\mu|)^2,
\end{equation}
are strictly greater than $0$ since $|\mu| \leq k < 1$. Since $a_{ij}$, $\alpha_1$, $\alpha_2$, and $\alpha_3$ do not explicitly depend on $x$ or $y$, Eq. (\ref{eq:elliptic_pdes}) can be written out as
\begin{multline}
\minus \text{div}(A\nabla u) = \minus \sum_{i=1}^2 \partial_i (A\nabla u)_i = \minus \sum_{i,k=1}^2  a_{ik}\partial_{ik}u.
\label{eq:elliptic_pde_real_part}
\end{multline}
 With that, we can define the two linear elliptic differential operators:
\begin{equation}
Lu := \minus \sum_{i,k=1}^2 a_{ik}\partial_{i,k}u,
\end{equation}
\begin{equation}
Lv := \minus \sum_{i,k=1}^2 a_{ik}\partial_{i,k}v.
\end{equation}
\noindent The analytical characteristics of the two differential operators are governed by the properties of $g$, such as adherence to the maximum principle. Similarly, the regularity of $g$ plays a critical role in determining the regularity of the associated lens mapping, particularly with respect to interior regularity. However, this work focuses on solving $Lu = 0$ and $Lv = 0$ numerically with appropriate boundary conditions.

\subsection{QCLens algorithm}
\renewcommand{\thealgorithm}{1}
  \begin{algorithm}[H]
    \caption{QCLens algorithm}
    \begin{algorithmic}[1]
      \Statex \textbf{Input:} plane domain \(\Omega\); map of the reduced shear \(g\); boundary conditions for real and imaginary part of the lens mapping \(f\) (Dirichlet, Neumann or mixed)
      \Statex \textbf{Output:} lens mapping $f$ (or deflection field $\beta$); map of convergence $\kappa$ and shear $\gamma$
      \State \(\mu(z) = \minus g(z) \quad \forall z \in \Omega\)
      \State Compute \(\alpha_1 = \frac{(\rho-1)^2 + \tau^2}{1-\rho^2-\tau^2};~\alpha_2 = \minus \frac{2\tau}{1-\rho^2-\tau^2};~\alpha_3 = \frac{(1+\rho)^2+\tau^2}{1-\rho^2-\tau^2} \quad \forall z \in \Omega \quad \) where \(\mu(z) = \rho(z) +\mathrm{i}\tau(z)\)
      \State Define the positive definite matrices A(z) := \(\Bigg(\begin{matrix}
                                                        \alpha_1(z) & \alpha_2(z)\\
                                                        \alpha_2(z) & \alpha_3(z)
                                                \end{matrix}\Bigg) \quad \forall z \in \Omega\)
      \For{\(w \in \{u=\mathrm{Re}(f),v=\mathrm{Im}(f)\}\)}
        \State Solve the elliptic PDE \(\minus \text{div}(A\nabla w) = 0\) on \(\Omega\)
        \EndFor
      \State \(\kappa = \frac{1}{2}(u_x + v_y)\); \(\gamma_1 = \frac{1}{2}(u_x - v_y)\) and \(\gamma_2 = \frac{1}{2}u_y\)
    \end{algorithmic}
   \label{algo:1}
  \end{algorithm}

\noindent The implementation of the QCLens algorithm utilizes the HiFlow3 software (\citealt{HiFlow3}), a C++-based multipurpose finite element solver. This approach discretizes the problem by employing a triangulation of the domain, $\Omega$, with the mesh width, $h$. The Beltrami coefficient, $- g$, is used to calculate the matrices, $A$, at each node. These matrices are integrated over the mesh using a two-dimensional quadrature formula to form the stiffness matrix and right-hand side vector. To solve the resulting elliptic PDEs for both the real and the imaginary parts of the lens mapping, finite element methods are employed. Specifically, piecewise linear functions are used to represent the solution in a finite-dimensional subspace. The solution is computed iteratively using a conjugate gradient solver, with Dirichlet or Neumann boundary conditions based on physical assumptions about the deflection field at the boundary $\partial \Omega$.

In the spirit of reproducible research, the QCLens algorithm is publicly available on GitHub\footnote{\url{https://github.com/JanJakob1/weak-lensing}}, including the material needed to reproduce the simulated experiments.

\subsection{Extension to the sphere}
Inversion methods for large areas of the sky, where the plane sky approximation can no longer be assumed, have become highly relevant with Stage IV surveys like \textit{Euclid} (\citealt{EuclidCollaboration2024}). Extending mass-mapping techniques to the sphere is a fundamental necessity for such surveys. 
A traditional approach is to decompose the sphere into overlapping patches, assume a flat approximation on each patch, reconstruct each patch independently, and recombine all patches on the sphere. It is natural to ask whether our results of the plane case can be generalized to the curved-sky treatment: If the plane sky approximation is considered as a coordinate chart around a given point on the curved surface, the Beltrami equation holds locally in this chart. However, the flat sky approximation does not provide isothermal coordinate charts, i.e., charts where the Riemannian metric is conformal to the Euclidean metric. To what extent the lens mapping can still be described in this setting as a q.c. mapping between curved surfaces will be the subject of future work.

\section{Experimental results}
\label{chap:5}

\subsection{Schwarzschild lens}
For the Schwarzschild lens, we obtain for the lens mapping
\begin{equation}
\begin{aligned}
f(z) = z - \frac{1}{\overline{z}} &= x\left( 1 - \frac{1}{x^2 + y^2} \right) 
+ i y \left( 1 - \frac{1}{x^2 + y^2} \right) \\
&=: u(x, y) + i v(x, y),
\label{eq:schwarzschild_lens}
\end{aligned}
\end{equation}
and for the Beltrami coefficient
\begin{equation}
\mu(z) = \frac{1}{\overline{z}^2} = \frac{x^2-y^2}{(x^2+y^2)^2} + \mathrm{i}\frac{2xy}{(x^2+y^2)^2} =: \rho(x,y) + \mathrm{i}\tau(x,y).
\end{equation} 

\begin{figure}[H]
    \centering
    \includegraphics[width=1.0\linewidth]{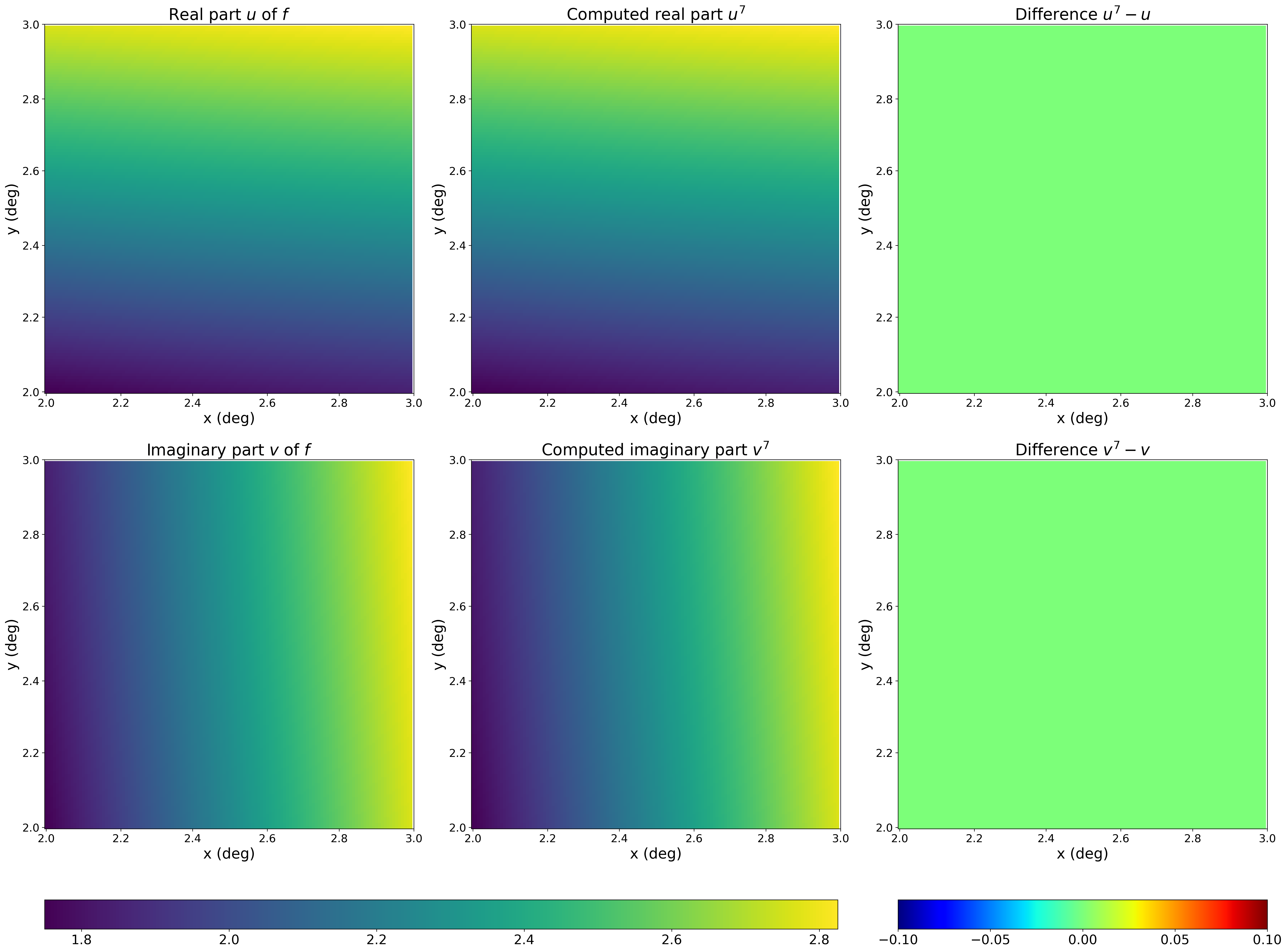}
    \caption{Schwarzschild lens: Comparison between actual lens mapping, $f = u+\mathrm{i}v$, and calculated lens mapping, $f^7 = u^7+\mathrm{i}v^7,$ with QCLens for a resolution of $n=7$ and Dirichlet boundary conditions.}
    \label{fig:exp1_lens_mapping_7}
\end{figure}

\noindent For both the real and imaginary part of $f$, we assumed Dirichlet boundary conditions. We computed $u^n$ and $v^n$ using the QCLens algorithm for different resolutions $n$ from $n=3$ to $n=8$ (i.e., $2^n$ calls per coordinate direction). As shown in Fig. \ref{fig:exp1_lens_mapping_7}, for $n=7$ we obtain an almost complete agreement between the actual and calculated lens mapping.

\begin{figure}[H]
    \centering
    \includegraphics[width=1.0\linewidth]{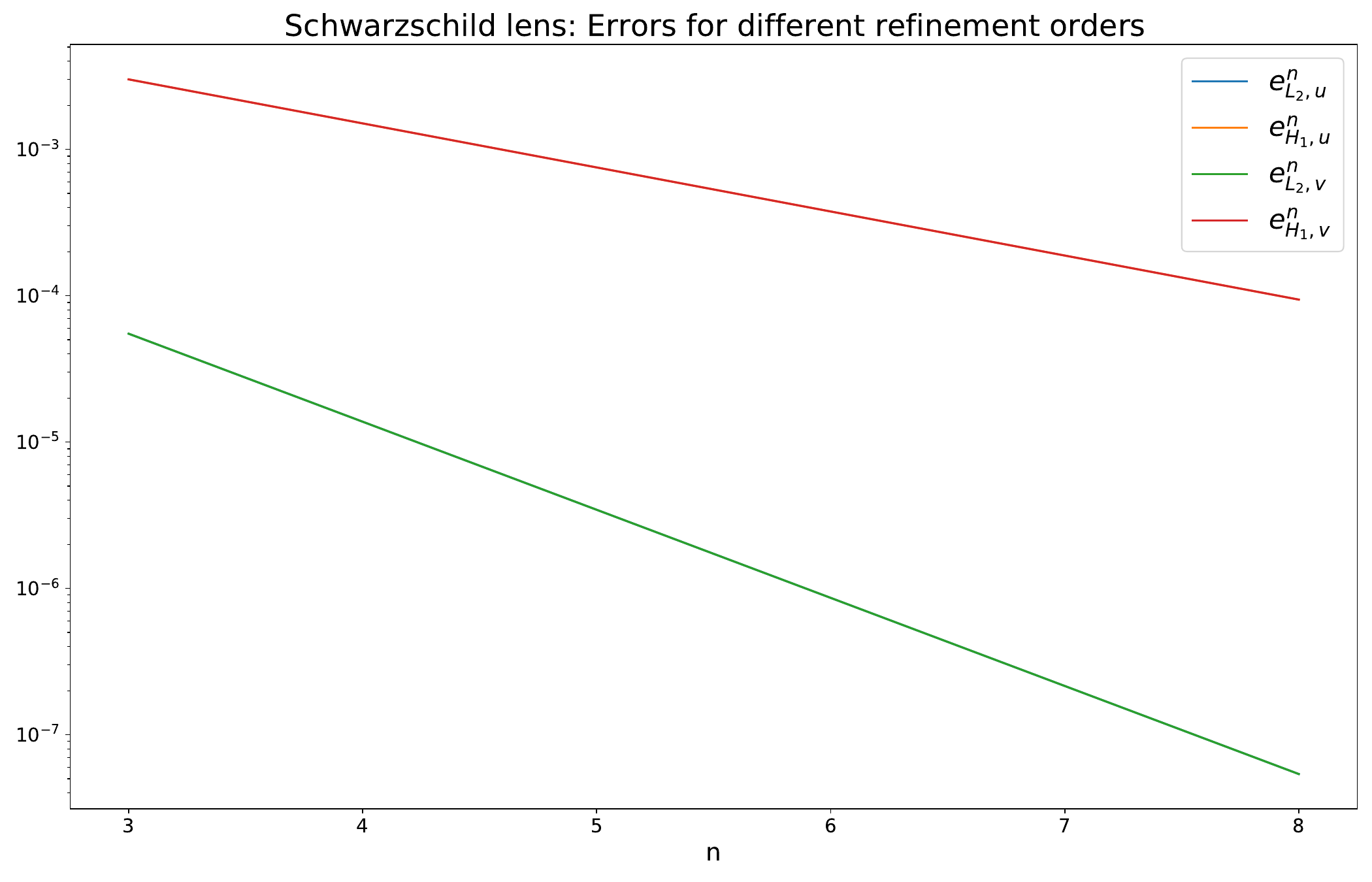}
    \caption{Schwarzschild lens error: $L_2$ and $H_1$ errors for different refinement orders with Dirichlet boundary conditions. The orange and red lines overlap, as do the green and blue lines.}
    \label{fig:errors_schwarzschild}
\end{figure}

\noindent Here, the coordinate system was chosen such that the point $(0,0)$ coincides with the position of the point mass in the lens plane. Additionally, the field of view, $\Omega$, is assumed to be the square $\{z=x+\mathrm{i}y~|~2 \leq x,y \leq 3\}$, such that we are in the weak lensing regime and do not cross any critical curve. As shown in Fig. \ref{fig:errors_schwarzschild}, the deviation between the real and the calculated lens mapping can also be quantified by plotting the $L_2$ and $H_1$ error against the refinement order, $n$, where
\begin{equation}
e_{L_2,w}^n = ||w-w^n||_{L_2} = \left(\int_{\Omega} \left(w(z)-w^n(z)\right)^2\mathrm{d}z \right)^{\frac{1}{2}},
\end{equation}
\begin{equation}
e_{H_1,w}^n = ||\nabla(w-w^n)||_{L_2} = \left(\int_{\Omega} |\nabla w(z)-\nabla w^n(z)|^2 \mathrm{d}z \right)^{\frac{1}{2}},
\end{equation}
with $w \in \{u,v\}$. The errors $e_{L_2,w}^n$ and $e_{H_1,w}^n$ are the same for $u$ and $v$.

\noindent As can be seen in Fig. \ref{fig:errors_schwarzschild}, the error $e_{L_2,w}^n$ decreases quadratically with increasing refinement order, $n$, while the error $e_{H_1,w}^n$ decreases linearly with $n$.

\subsection{Singular isothermal lens}
\begin{figure}[H]
    \centering
    \includegraphics[width=0.90\linewidth]{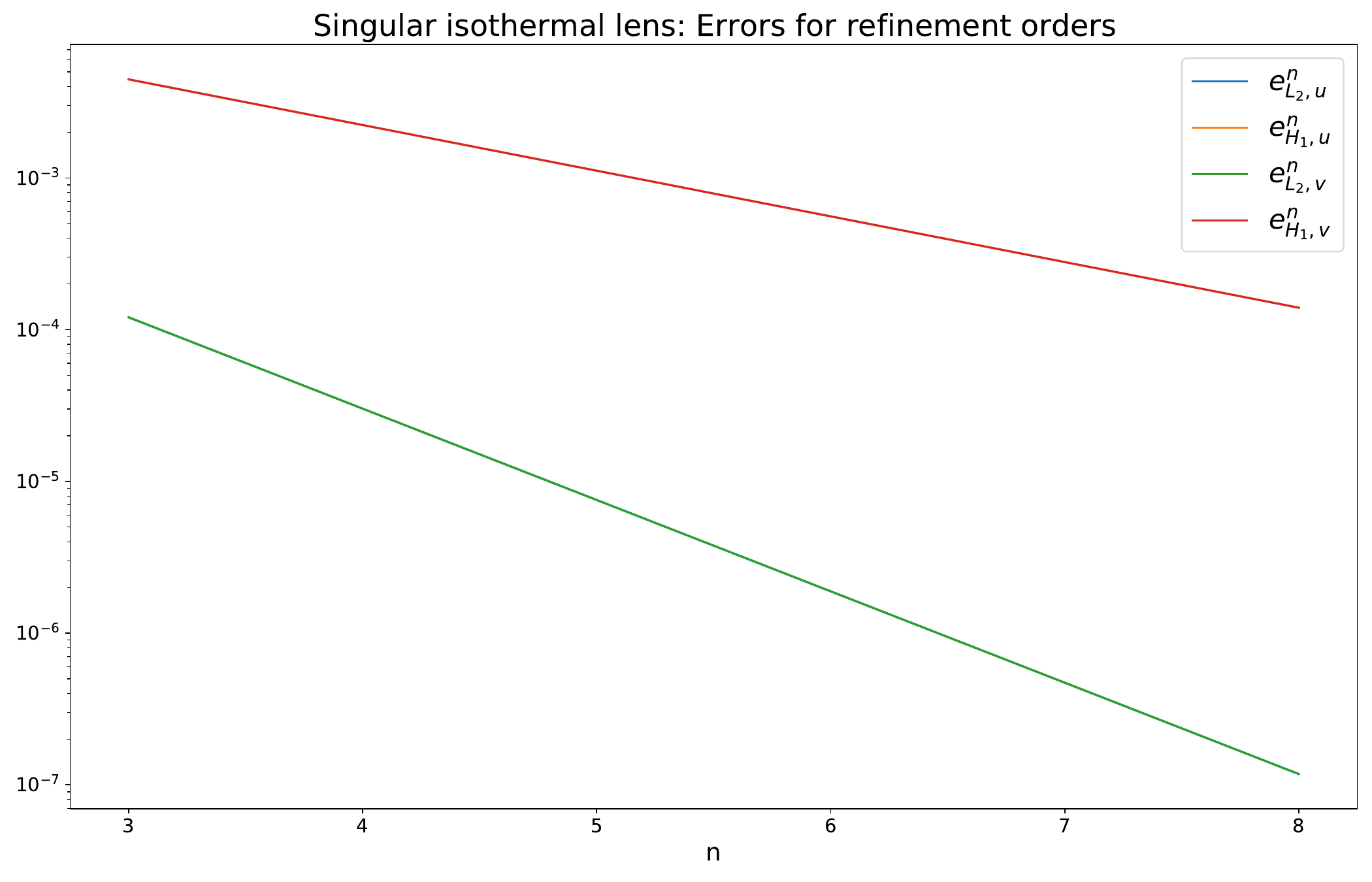}
    \caption{Singular isothermal lens error: $L_2$ and $H_1$ errors for different refinement orders with Dirichlet boundary conditions. The orange   and red lines overlap, as do the green and blue lines.}
    \label{fig:errors_isothermal}
\end{figure}

For the singular isothermal lens, we obtain for the lens mapping
\begin{equation}
\begin{aligned}
f(z) = z - \frac{z}{|\overline{z}|} &= x\left(1-\frac{1}{\sqrt{x^2+y^2}}\right) + \mathrm{i}y\left(1-\frac{1}{\sqrt{x^2+y^2}}\right)\\ &=: u(x,y) + \mathrm{i}v(x,y),
\end{aligned}
\end{equation}
and for the Beltrami coefficient
\begingroup
\small
\begin{equation}
\begin{aligned}
\mu(z) = \frac{z^2}{2|\overline{z}|^3 - |\overline{z}|^2} &= \frac{\left(x^{2} - y^{2}\right)}{2 \left(x^{2} + y^{2}\right)^{\frac{3}{2}} - \left(x^{2} + y^{2}\right)} + \mathrm{i}\frac{2xy}{2 \left(x^{2} + y^{2}\right)^{\frac{3}{2}} - \left(x^{2} + y^{2}\right)}\\ &=: \rho(x,y) + \mathrm{i}\tau(x,y).
\end{aligned}
\end{equation}
\endgroup
\begin{figure}[H]
    \centering
    \includegraphics[width=1.0\linewidth]{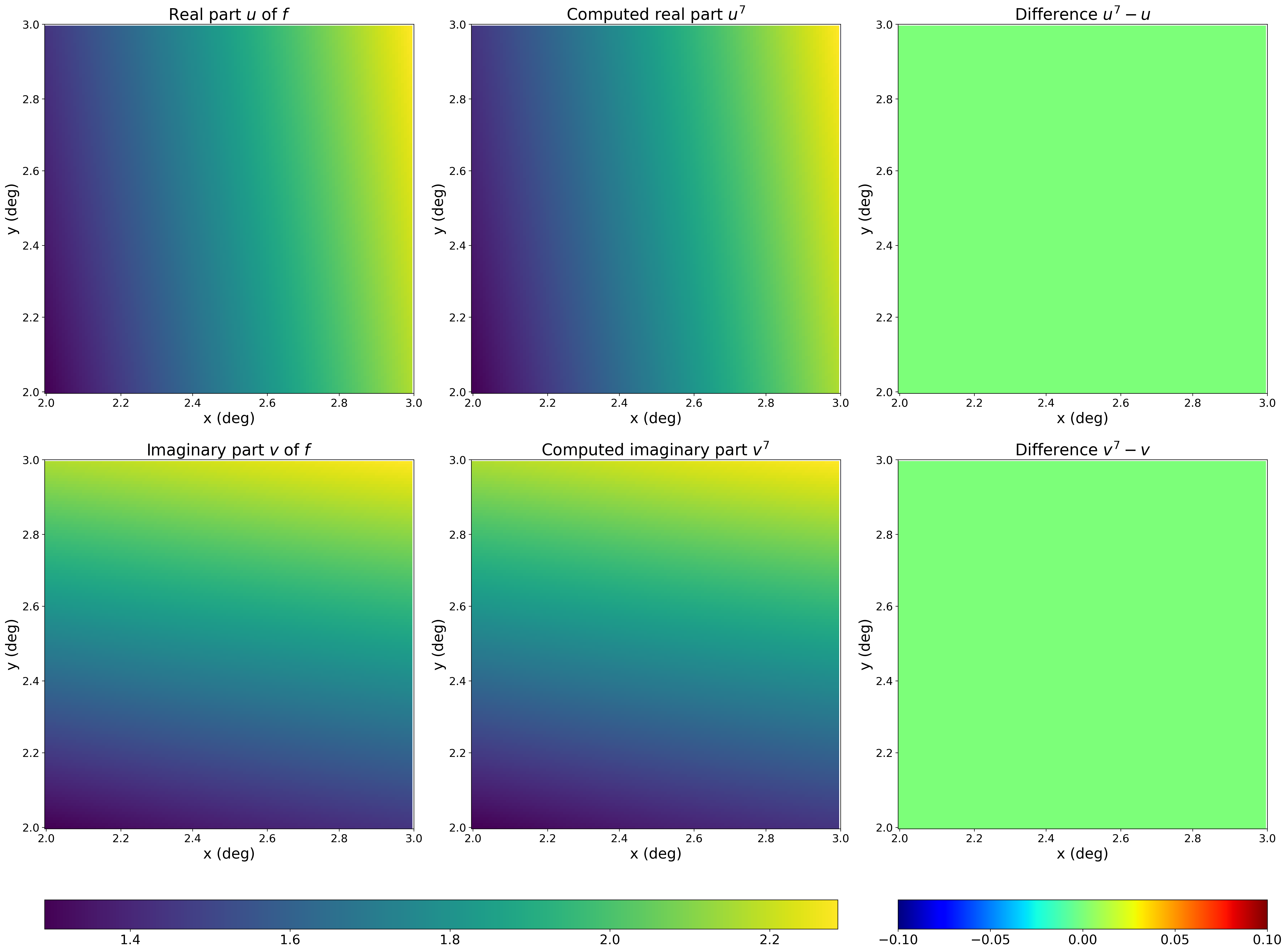}
    \caption{Singular isothermal lens: Comparison between actual lens mapping, $f = u+\mathrm{i}v$, and calculated lens mapping, $f^7 = u^7+\mathrm{i}v^7$, with QCLens for a resolution of $n=7$ and Dirichlet boundary conditions.}
    \label{fig:exp2_lens_mapping_7}
\end{figure}
\noindent Using the same coordinate system and field of view as for the Schwarzschild lens, we obtain analogous results under the assumption of Dirichlet boundary conditions. As in the Schwarzschild lens experiment, the lens mapping computed for $n=7$ shows near-perfect agreement with the actual mapping, as illustrated in Fig. \ref{fig:exp2_lens_mapping_7}. Correspondingly, Fig. \ref{fig:errors_isothermal} demonstrates that the error $e_{L_2,w}^n$ decreases quadratically with increasing refinement order $n$, while the error $e_{H_1,w}^n$ exhibits a linear decrease.

\section{Conclusion}

We have proposed a novel inversion algorithm for weak gravitational lensing based on the q.c. mapping framework. By reformulating the lens equation as a Beltrami equation, the problem was reduced to solving elliptic PDEs for the real and the imaginary part of the lens mapping. The QCLens algorithm was applied to analytically solvable cases, such as the Schwarzschild and singular isothermal lens models, demonstrating consistency with expected results. Additionally, we have discussed the feasibility of extending the approach to spherical geometries, which will be necessary for future surveys like \textit{Euclid}. These findings provide a foundation for further exploration of mass-mapping techniques within this framework.

\bibliographystyle{aa} 
\bibliography{refs}

\end{document}